\documentclass[prl,twocolumn,superscriptaddress]{revtex4-1}
\usepackage{}
\usepackage{amssymb}
\usepackage{amsfonts}
\usepackage{graphics,graphicx,epsfig,bm,amsmath,amsthm,amssymb}
\usepackage{bm}
\usepackage{bbm}
\usepackage{longtable}
\usepackage{multirow}
\usepackage{array}
\usepackage{color}
\usepackage[usenames,dvipsnames]{xcolor}

\usepackage{float}
\usepackage[a4paper,colorlinks=true,
linkcolor=blue,citecolor=blue,
pdfauthor={ },
pdftitle={ },
pdfsubject={ },
pdfkeywords={ }]{hyperref}

\begin{document}

%\title{Experimental Demonstration of Triple Uncertainty Relations for Spins}
%\title{Experimental Demonstration of Triple Uncertainty Relations for Angular Momentum}
\title{Experimental Demonstration of Uncertainty Relations\\for the Triple Components of Angular Momentum}

\author{Wenchao Ma}
%\email{These authors contributed equally to this work.}
\affiliation{CAS Key Laboratory of Microscale Magnetic Resonance and Department of Modern Physics, University of Science and Technology of China, Hefei 230026, China}

\author{Bin Chen}
%\email{These authors contributed equally to this work.}
\affiliation{State Key Laboratory of Low-Dimensional Quantum Physics and Department of Physics, Tsinghua University, Beijing 100084, China}
\affiliation{Tsinghua National Laboratory for Information Science and Technology, Beijing 100084, China}

\author{Ying Liu}
\affiliation{CAS Key Laboratory of Microscale Magnetic Resonance and Department of Modern Physics, University of Science and Technology of China, Hefei 230026, China}

\author{Mengqi Wang}
\affiliation{CAS Key Laboratory of Microscale Magnetic Resonance and Department of Modern Physics, University of Science and Technology of China, Hefei 230026, China}

\author{Xiangyu Ye}
\affiliation{CAS Key Laboratory of Microscale Magnetic Resonance and Department of Modern Physics, University of Science and Technology of China, Hefei 230026, China}

\author{Fei Kong}
\affiliation{CAS Key Laboratory of Microscale Magnetic Resonance and Department of Modern Physics, University of Science and Technology of China, Hefei 230026, China}

\author{Fazhan Shi}
\affiliation{CAS Key Laboratory of Microscale Magnetic Resonance and Department of Modern Physics, University of Science and Technology of China, Hefei 230026, China}
\affiliation{Synergetic Innovation Center of Quantum Information and Quantum Physics, University of Science and Technology of China, Hefei, Anhui 230026, China}

\author{Shao-Ming Fei}
\email{feishm@cnu.edu.cn}
\affiliation{School of Mathematics and Sciences, Capital Normal University, Beijing 100048, China}

\author{Jiangfeng Du}
\email{djf@ustc.edu.cn}
\affiliation{CAS Key Laboratory of Microscale Magnetic Resonance and Department of Modern Physics, University of Science and Technology of China, Hefei 230026, China}
\affiliation{Synergetic Innovation Center of Quantum Information and Quantum Physics, University of Science and Technology of China, Hefei, Anhui 230026, China}

%\pacs{76.30.Mi (color centers), 76.70.Hb (ODMR), 07.55.Ge (Magnetometry)}

\begin{abstract}
The uncertainty principle is considered to be one of the most striking features in quantum mechanics.
In the textbook literature, uncertainty relations usually refer to the preparation uncertainty which imposes a limitation on the spread of measurement outcomes for a pair of non-commuting observables.
In this work, we study the preparation uncertainty for the angular momentum, especially for spin-$1/2$. We derive uncertainty relations encompassing the triple components of angular momentum, and show that compared with the relations involving only two components, a triple constant $2/\sqrt3$ often arises. Intriguingly, this constant is the same for the position and momentum case.
Experimental verification is carried out on a single spin in diamond, and the results confirm the triple constant in a wide range of experimental parameters.

%We study the uncertainty relations for the triple components of spin-$1/2$ operators. Both standard deviation-based product form uncertainty relations and summation form uncertainty relations of variance-based uncertainty relations for three Pauli matrices are investigated.
%Tight uncertainty relations are obtained from a universal triple constant $\tau$.
%Experimental verification of these uncertainty relations are carried out on a single spin in diamond.
\end{abstract}
\maketitle

\emph{Introduction}.---
The uncertainty principle was first proposed by Heisenberg in a thought experiment showing that the measurement of an electron's position disturbs the momentum inevitably \cite{Heisenberg}. In the ensuing few years, Kennard \cite{Kennard},
Weyl \cite{Weyl}, Robertson \cite{Robertson29}, and Schr\"{o}dinger \cite{Schroedinger} derived mathematically rigorous relations, such as the famous Heisenberg-Robertson uncertainty relation \cite{Robertson29}
\begin{equation}\label{UR}
\Delta A\Delta B\geq\frac{1}{2}|\langle[A,B]\rangle|,
\end{equation}
with the standard deviation $\Delta\Omega=\sqrt{\langle \Omega^{2}\rangle-\langle \Omega\rangle^{2}}$ for the observable $\Omega=A$ or $B$, the angle brackets $\langle\rangle$ denoting the expectation of an operator with respect to a given state $\rho$, and $[A,B]=AB-BA$.
The inequality imposes a tradeoff between the statistical dispersions $\Delta A$ and $\Delta B$ of the pair of non-commuting observables $A$ and $B$ for the given quantum state $\rho$.
%The inequality imposes a restriction on the product of $\Delta A$ and $\Delta B$, thus presenting a trade-off between the statistical dispersions of a pair of non-commuting observables for a given quantum state.
This type of uncertainty, often termed as preparation uncertainty, deals with the spread of measurement outcomes rather than Heisenberg's original idea which investigates measurement inaccuracies \cite{Busch07,Busch13PRL,Busch14PRA,Busch14RMP,Buscemi,Sulyok,Ma}. In this Letter, we only discuss the preparation uncertainty.

%This kind of inequalities deal with the so-called preparation uncertainty which states that one cannot prepare a state $\rho$ so that $\Delta A$ and $\Delta B$ simultaneously become arbitrarily small for non-commuting observables $A$ and $B$.

%Uncertainty principle is considered to be one of the most striking features in quantum mechanics. The corresponding uncertainty relations imply the impossibility of simultaneously determining the definite values of non-commuting observables. Based on the measurement outcomes, such uncertainty relations can be described in various ways \cite{y6,BLW1,y7,y8,y16,y17,y9,y10,y11,y12,y13,y14,Busch,Maccone,Chen,Busch13PRL,Busch14PRA,Busch14RMP,BLW}.
%The famous Heisenberg-Robertson uncertainty relation is the typical variance-based \cite{Heisenberg,Robertson29},
%%Mathematically, this can be expressed by the well-known Heisenberg-Robertson uncertainty relation :
%\begin{equation}\label{UR}
%\Delta A\Delta B\geq\frac{1}{2}|\langle[A,B]\rangle|,
%\end{equation}
%where the standard deviation $\Delta\Omega=\sqrt{\langle \Omega^{2}\rangle-\langle \Omega\rangle^{2}}$, $\langle \Omega\rangle$ denotes the expectation of the observable $\Omega$, $[A,B]=AB-BA$.
%The Heisenberg-Robertson uncertainty relation
%presents a lower bound on the product of the standard deviations of two observables, and
%provides a trade-off relation of measurement errors of these two observables for any given quantum states.

Uncertainty principle can be described by various relations \cite{Hirschman,Beckner,Birula75,Deutsch,Kraus,Maassen,Braunstein,Sanchez98,Ghirardi,Birula06,Wehner10,Birula11,Partovi,Puchala,Friedland,Maccone,Zhang,Li,Coles}, but most well-known uncertainty relations deal with two observables till now. %till now we have no product form uncertainty relations for more than two observables.
In contrast to the two-observable relation in (\ref{UR}) which can be derived via the Cauchy-Schwarz inequality, it is difficult to obtain a nontrivial multi-observable uncertainty relation that has a form similar to (\ref{UR}), although attempts to encompass three or more observables have a long history since Robertson's work in 1934 \cite{Robertson34,Ivanovic,Sanchez93,Sanchez95,Trifonov,Shirokov,Pati,Wehner08,Huang,Kaniewski,Chen,Abbott,Chen2016-1,Chen2016-2}.
%As there is no relations like Schwartz inequality for three or more objects, it is difficult to have
%a nontrivial inequality satisfied by the quantity $(\Delta A)^{2}(\Delta B)^{2}...(\Delta C)^{2}$ generally, although attempts to encompass three or more observables have a long history \cite{Robertson34,Trifonov,Shirokov}.
Recently, an uncertainty relation for three pairwise canonical observables $p,q,r$ satisfying $[p,q]=[q,r]=[r,p]=-i\hbar$ with $r=-p-q$ was derived by Kechrimparis and Weigert \cite{Weigert}.
Here ${p}$ and ${q}$ are the momentum and position, respectively.
From the inequality (\ref{UR}) one immediately obtains
$\Delta p\Delta q \ge {\hbar }/{2}$, $\Delta q\Delta r \ge {\hbar }/{2}$, $\Delta r\Delta p \ge {\hbar }/{2}$. By multiplying these inequalities and taking the square root, one gets
$\Delta p\Delta q\Delta r\geq({\hbar}/{2})^{\frac{3}{2}}$.
However, this inequality is not tight. In other words, there is no state satisfying such lower bound. By introducing the triple constant $\tau=2/\sqrt{3}$, the tight triple uncertainty relation $\Delta p\Delta q\Delta r \geq(\tau{\hbar}/{2})^{\frac{3}{2}}$ is established.
%The `observable' ${r}$ is not a physical quantity, neither independent in this triple.

In contrast to the couple of observables $p$ and $q$, the latecomer $r$ seems artificial and may not exhibit an explicit physical meaning.
Yet the three components of angular momentum form a natural triple \cite{Dammeier}.
%In fact, besides the dual observables like position and momentum, there are also triple physical observables like the angular momentum.
In this Letter, we formulate tight uncertainty relations satisfied by the triple components of angular momentum, and show that the triple constant $\tau$ also arises.
%we focus on the uncertainty relations satisfied by the angular momentum and formulate tight triple uncertainty relations where the triple constant $\tau$ also arises.
An experimental test is performed on a single spin in diamond.
%An experimental test is performed on a negatively charged nitrogen-vacancy (NV) in diamond.

\emph{Uncertainty relations}.---
In this Letter, we always set $\hbar=1$.
Let $S_x$, $S_y$, and $S_z$ be the angular momentum operators satisfying commutation relations
$[S_x,S_y]=i S_z$, $[S_z,S_x]=i S_y$, and $[S_y,S_z]=i S_x$. From the inequality (\ref{UR}) one obtains
\begin{equation}\label{spin-pair-product}
\begin{aligned}
&\Delta {S_x} \Delta {S_y} \ge \frac{1}{2}\left| {\left\langle {{S_z}} \right\rangle } \right|,\\
&\Delta {S_y} \Delta {S_z} \ge \frac{1}{2}\left| {\left\langle {{S_x}} \right\rangle } \right|,\\
&\Delta {S_z} \Delta {S_x} \ge \frac{1}{2}\left| {\left\langle {{S_y}} \right\rangle } \right|.
\end{aligned}
\end{equation}
By multiplying the above inequalities and taking the square root, one directly gets a trivial relation
$\Delta {S_x}  \Delta {S_y}  \Delta {S_z} \ge  {\left| {\langle {S_x}\rangle }
{\langle {S_y}\rangle }  {\langle {S_z}\rangle }/8 \right|}^{\frac{1}{2}}$,
where the equality holds only when both sides are zero.
%Similar to the inequality (\ref{pqr-trivial}), the lower bound in the inequality (\ref{spin-trivial-product}) cannot be attained.
We tighten the lower bound for spin-$1/2$ by introducing the triple constant $\tau$ (see sketch of proof in Supplemental Material \cite{sm}), i.e.,
%The following is our main result on nontrivial triple uncertainty relation in product form,
\begin{equation}\label{spin-product}
\Delta {S_x}\Delta {S_y}\Delta {S_z} \ge {\left| {\frac{{{\tau ^3}}}{8}\langle {S_x}\rangle \langle {S_y}\rangle \langle {S_z}\rangle } \right|^{\frac{1}{2}}}.
\end{equation}
The equality in (\ref{spin-product}) holds when $\left| {{r_x}} \right| = \left| {{r_y}} \right| = \left| {{r_z}} \right| = 1/\sqrt 3$ or $\left( {\left| {{r_x}} \right| - 1} \right)\left( {\left| {{r_y}} \right| - 1} \right)\left( {\left| {{r_z}} \right| - 1} \right) = 0$. Here $r_x,r_y,r_z$ are the components of the Bloch vector $\bm{r}$ of a qubit state with the density matrix $\rho  = ({\mathbbm{1} + {\bm{r}} \cdot {\bm{\sigma }}})/2$.

\begin{figure}
\includegraphics[width=0.8\columnwidth]{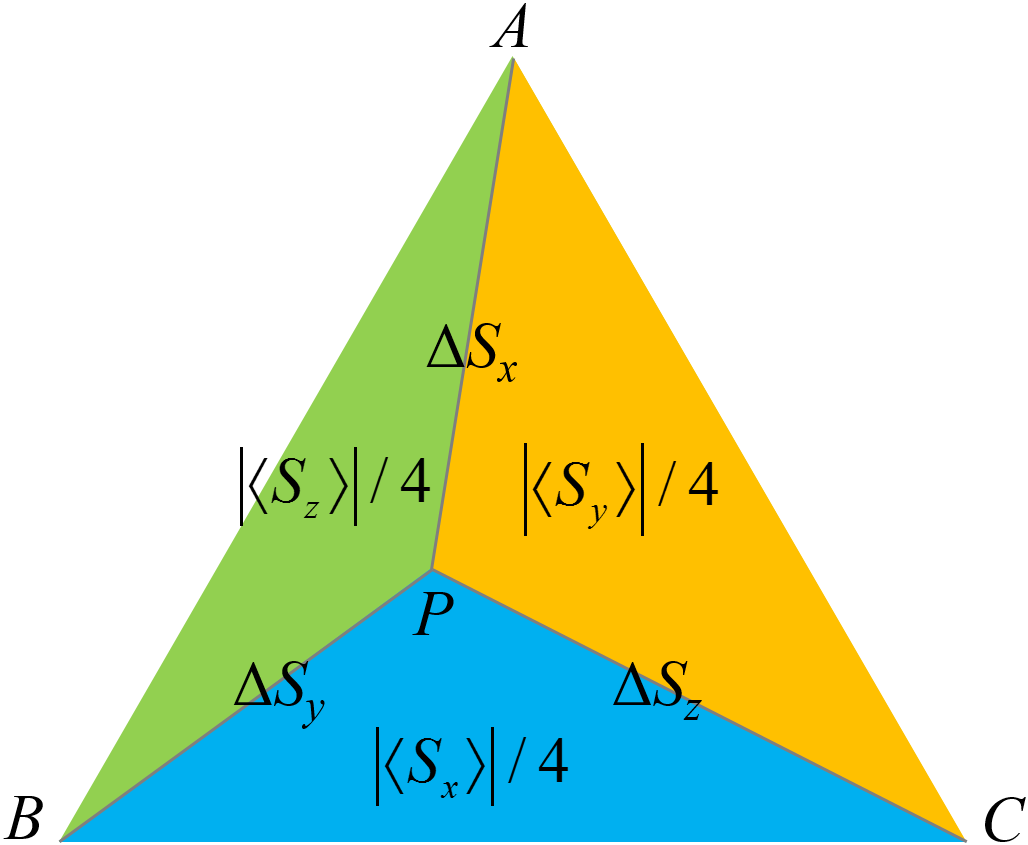}
\caption{Geometric analog of uncertainty relations (\ref{spin-pair-product}), (\ref{spin-product}), (\ref{spin-pair-sum}), and (\ref{spin-sum}).
   The equilateral triangle has vertices $A,B,C$ and a point $P$ inside. The lengths of the line segments $PA$, $PB$, and $PC$ are denoted by $|PA|$, $|PB|$, and $|PC|$, respectively. The areas of the triangles $PAB$, $PBC$, and $PCA$ are denoted by $|\triangle PAB|$, $|\triangle PBC|$, and $|\triangle PCA|$, respectively. These geometric quantities have the same relations as the inequalities (\ref{spin-pair-product}), (\ref{spin-product}), (\ref{spin-pair-sum}), and (\ref{spin-sum}) under the correspondence $|PA|\leftrightarrow\Delta S_x$, $|PB|\leftrightarrow\Delta S_y$, $|PC|\leftrightarrow\Delta S_z$ and $|\triangle PAB|\leftrightarrow|\langle {S_z}\rangle|/4$, $|\triangle PBC|\leftrightarrow|\langle {S_x}\rangle|/4$, $|\triangle PCA|\leftrightarrow|\langle {S_y}\rangle|/4$.
   }
   %Set $|PA|=\Delta S_x$, $|PB|=\Delta S_y$, $|PC|=\Delta S_z$ and $|\triangle PAB|=|\langle {S_z}\rangle|/4$, $|\triangle PBC|=|\langle {S_x}\rangle|/4$, $|\triangle PCA|=|\langle {S_y}\rangle|/4$. Then these geometric quantities give rise to inequalities (\ref{spin-pair-product}), (\ref{spin-product}), (\ref{spin-pair-sum}), and (\ref{spin-sum}).
    \label{triangle}
\end{figure}

Besides multiplicative form uncertainty relations, one may also tighten additive form relations.
The inequalities in (\ref{spin-pair-product}) entail
%The AM-GM (?) inequality and the inequalities in (\ref{spin-pair-product}) entail the following relations,
\begin{equation}\label{spin-pair-sum}
\begin{aligned}
 {\left( {\Delta {S_x}} \right)^2} + {\left( {\Delta {S_y}} \right)^2} \ge |\langle {S_z}\rangle|,  \\
 {\left( {\Delta {S_y}} \right)^2} + {\left( {\Delta {S_z}} \right)^2} \ge |\langle {S_x}\rangle|, \\
 {\left( {\Delta {S_z}} \right)^2} + {\left( {\Delta {S_x}} \right)^2} \ge |\langle {S_y}\rangle|. \\
\end{aligned}
\end{equation}
From the above inequalities one immediately gets ${\left( {\Delta {S_x}} \right)^2} + {\left( {\Delta {S_y}} \right)^2} + {\left( {\Delta {S_z}} \right)^2} \ge \left( {\left| {\langle {S_x}\rangle } \right| + \left| {\langle {S_y}\rangle } \right| + \left| {\langle {S_z}\rangle } \right|} \right)/2$,
%$$
%\begin{aligned}
%&{\left( {\Delta {S_x}} \right)^2} + {\left( {\Delta {S_y}} \right)^2} + {\left( {\Delta {S_z}} \right)^2} \ge \\
%&\frac{1}{2}\left( {\left| {\langle {S_x}\rangle } \right| + \left| {\langle {S_y}\rangle } \right| + \left| {\langle {S_z}\rangle } \right|} \right),
%\end{aligned}
%$$
which is again not tight. We also tighten the lower bound by introducing the triple constant $\tau$ (see sketch of proof in Supplemental Material \cite{sm}), i.e.,
%a tight lower bound requires the triple constant $\tau$, i.e.,
%The tight triple uncertainty relation in the summation form has a form, see Supplemental Material for proof,
\begin{equation}\label{spin-sum}
\begin{aligned}
&{\left( {\Delta {S_x}} \right)^2} + {\left( {\Delta {S_y}} \right)^2} + {\left( {\Delta {S_z}} \right)^2} \ge \\
&\frac{\tau}{2}\left( {\left| {\langle {S_x}\rangle } \right| + \left| {\langle {S_y}\rangle } \right| + \left| {\langle {S_z}\rangle } \right|} \right).
\end{aligned}
\end{equation}
The equality in (\ref{spin-sum}) is attained if and only if $\left| {{r_x}} \right| = \left| {{r_y}} \right| = \left| {{r_z}} \right| = 1/\sqrt 3$ for spin-$1/2$.
%Here we have also provided a tight trade-off relation between sum of absolute mean values and sum of variances of $S_{x},S_{y},S_{z}$.
%If the sum uncertainties of $S_{x},S_{y},S_{z}$ is fixed, for instance, the measured state is pure, then the value of $|\langle S_{x}\rangle|+|\langle S_{y}\rangle|+|\langle S_{z}\rangle|$ is subjected to this tight upper bound.
%In fact, the inequality (\ref{spin-sum}) is valid for spin-$s$ with $s=1/2, ~1, ~3/2, ~2, ~\cdots\cdots$.

Interestingly, the uncertainty relations (\ref{spin-pair-product}), (\ref{spin-product}), (\ref{spin-pair-sum}), and (\ref{spin-sum}) are analogous to the geometric relations of an equilateral triangle, as depicted in Fig.~\ref{triangle} (see proofs of these geometric relations in Supplemental Material \cite{sm}). We leave the experimental demonstrations of the uncertainty relations (\ref{spin-product}) and (\ref{spin-sum}) in Section Experiment.

The uncertainty relations (\ref{spin-product}) and (\ref{spin-sum}) have state-dependent lower bounds. It is similar for uncertainty relations with state-independent lower bounds.
The pairwise inequalities for spin-$1/2$ \cite{Busch14PRA}, namely,
%The uncertainty relation (\ref{spin-sum}) is state-dependent. It is also to derive
%uncertainty relations with state-independent lower bounds. In fact, one has the following pairwise relations for $S_x$, $S_y$ and $S_z$,
$ {\left( {\Delta {S_x}} \right)^2} + {\left( {\Delta {S_y}} \right)^2} \ge {1}/{4}$,
$ {\left( {\Delta {S_y}} \right)^2} + {\left( {\Delta {S_z}} \right)^2} \ge {1}/{4}$, and
$ {\left( {\Delta {S_z}} \right)^2} + {\left( {\Delta {S_x}} \right)^2} \ge {1}/{4}$,
immediately yield the inequality
%which yield an inequality that is not tight,
${\left( {\Delta {S_x}} \right)^2} + {\left( {\Delta {S_y}} \right)^2} + {\left( {\Delta {S_z}} \right)^2} \ge {3}/{8}$ which is not tight.
A tight lower bound needs an additional factor ${\tau}^2$ \cite{Abbott,Hofmann}, i.e.,
%From our approach we have the following tight relation,
\begin{equation}\label{spin-sum-1/2}
{\left( {\Delta {S_x}} \right)^2} + {\left( {\Delta {S_y}} \right)^2} + {\left( {\Delta {S_z}} \right)^2} \ge \frac{1}{2} = \frac{3}{8}{\tau}^2.
\end{equation}
The equality is attained if and only if $\left| {\bm{r}} \right| = 1$, i.e., the qubit is in a pure state. The relation (\ref{spin-sum-1/2}) is also supported by our experiment.

Here it should be noted that the inequality (\ref{spin-sum}) is in fact valid for any spin quantum number.
The relation (\ref{spin-sum-1/2}) turns into
\begin{equation}\label{prominent}
{\left( {\Delta {S_x}} \right)^2} + {\left( {\Delta {S_y}} \right)^2} + {\left( {\Delta {S_z}} \right)^2} \ge s
\end{equation}
for the spin quantum number $s$ \cite{Hofmann}, and the factor ${\tau}^2$ does not hold for $s \geq 1$ (see explanation in Supplemental Material \cite{sm}).
%However, the state-independent relation (\ref{spin-sum-1/2}) does not hold for $s \geq 1$.
It should also be noted that although the state-dependent lower bound in the inequality (\ref{spin-product}) vanishes in some cases, this uncertainty relation is not covered by the prominent relation (\ref{prominent}) and is valuable in its own right.

\begin{figure}
\includegraphics[width=0.9\columnwidth]{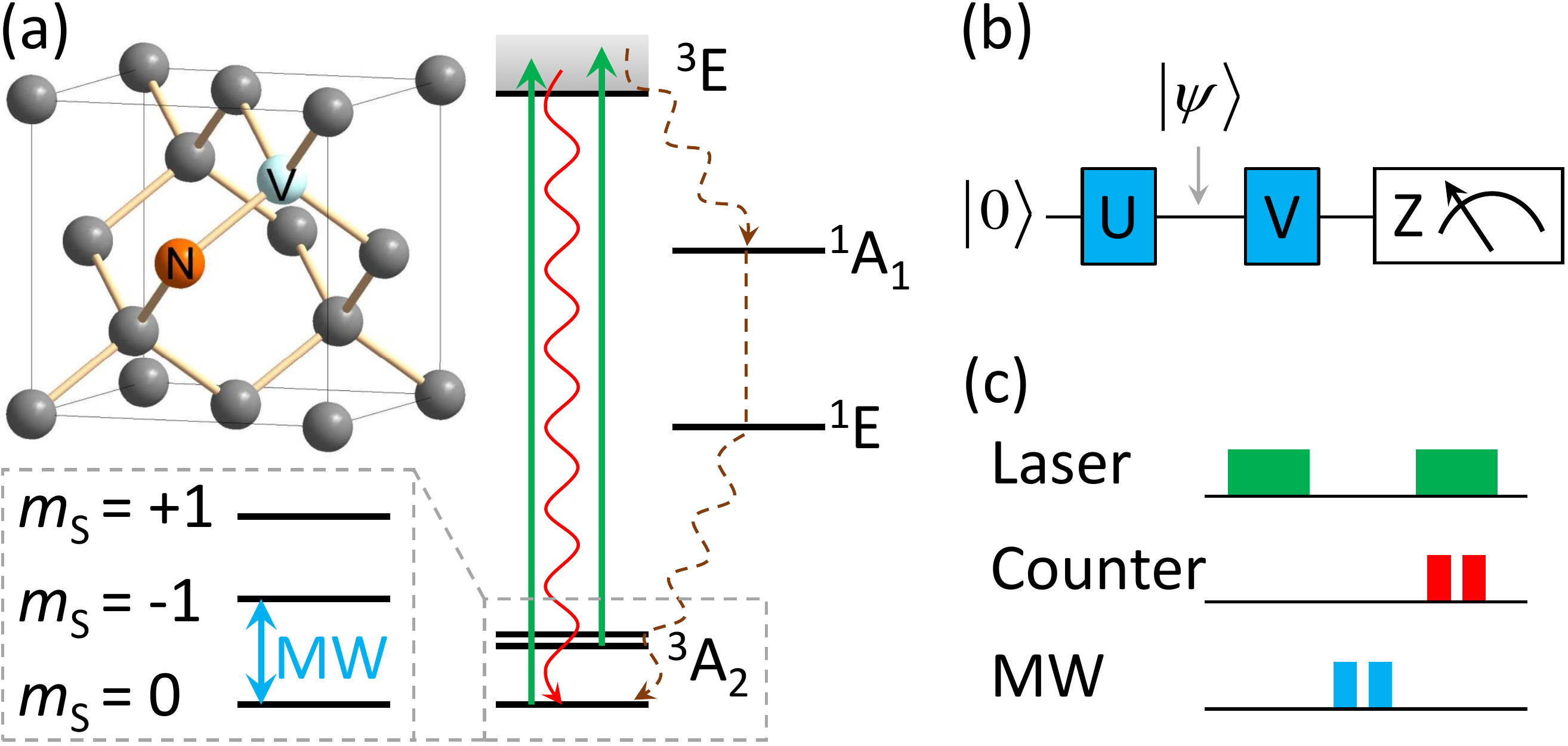}
\caption{Experimental system and method.
   (a) Negatively charged NV center in diamond and the electronic energy level structure. The NV center consists of a substitutional nitrogen atom and a neighboring vacancy. The electronic ground state $^3A_2$ is a triplet state, where two levels are encoded as a qubit and exploited in the experiment.
   (b) Quantum circuit for qubit control and measurement.
   (c) Pulse sequence for implementing the quantum circuit. In the experiment, such process is repeated four million times.
   }
    \label{structure}
\end{figure}
\emph{Experimental demonstration}.---
To verify the uncertainty relations (\ref{spin-product}) and (\ref{spin-sum}), we carry out the experiment on a negatively charged nitrogen-vacancy (NV) center in diamond.
The single spins of NV centers are convenient to initialize and read out, have long coherence time, and can be manipulated with high precision. These advantages enable NV centers to be widely applied in nanoscale sensing, quantum information, and fundamental physics \cite{Doherty,Schirhagl,Prawer,Wrachtrup}.
%quantum metrology \cite{Waldherr2012,Liu}

\begin{figure}\centering
\includegraphics[width=1\columnwidth]{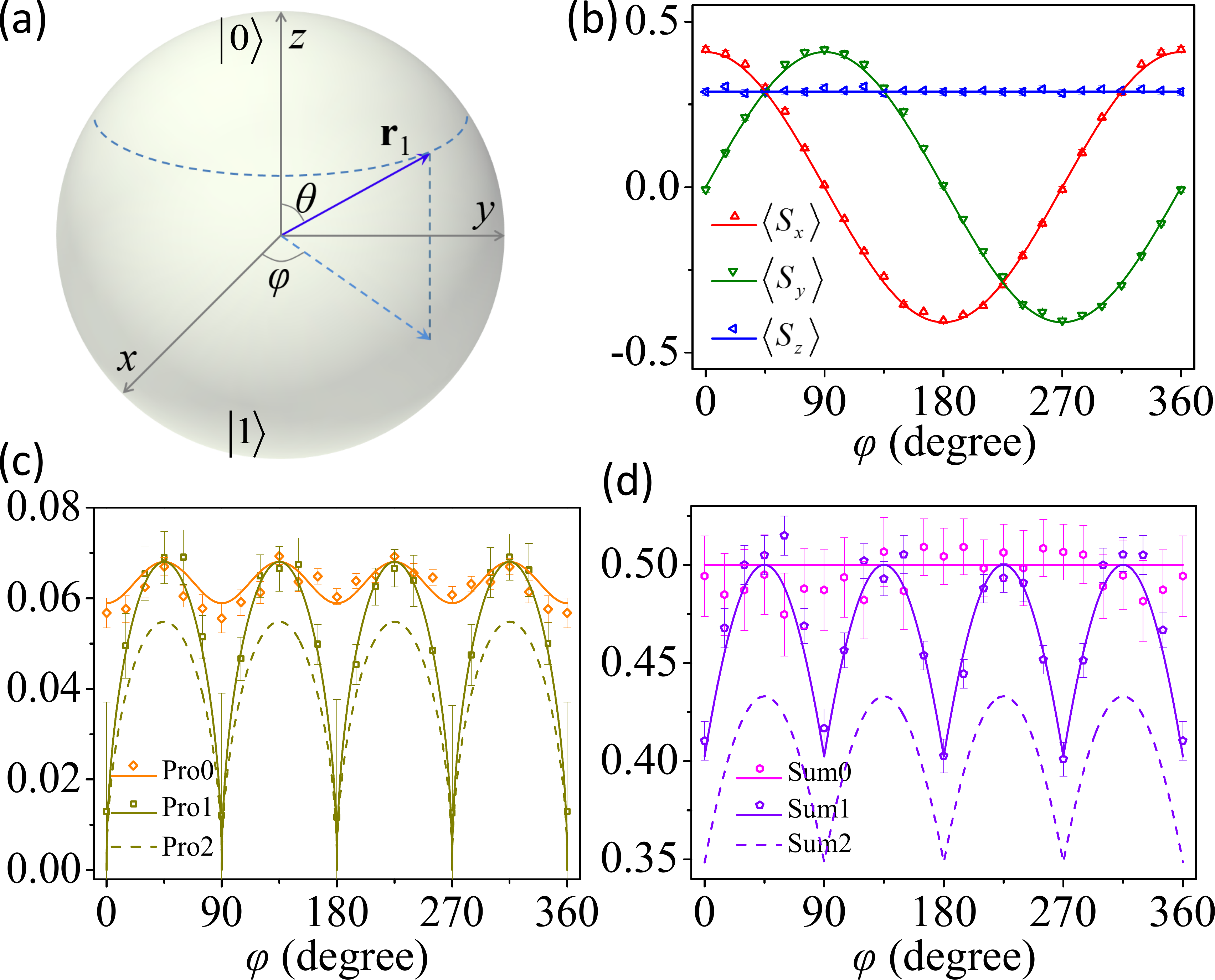}
\caption{Experimental results.
   (a) Bloch sphere and the vector ${\bf{r}}_1$ with $\theta  = \arctan \sqrt 2$.
   (b) Expectations of $S_x$, $S_y$, and $S_z$. The red, olive, and blue curves, in turn, represent the theoretical values of $\langle {S_x}\rangle$, $\langle {S_y}\rangle$, and $\langle {S_z}\rangle$. The corresponding scattered points represent the experimental values.
    %Some error bars are shorter than the labels of the data points.
   (c) Experimental demonstration for the uncertainty relation of the multiplicative form. The solid orange, solid green, and dashed green curves, in turn, represent the theoretical values of Pro0, Pro1, and Pro2, where $\textrm{Pro0}=\Delta {S_x} \Delta {S_y} \Delta {S_z}$, $\textrm{Pro1}=\left|\tau ^3 {\langle {S_x}\rangle } {\langle {S_y}\rangle } {\langle {S_z}\rangle }/8 \right|^{1/2}$, and $\textrm{Pro2}=\left|{\langle {S_x}\rangle } {\langle {S_y}\rangle } {\langle {S_z}\rangle }/8 \right|^{1/2}$. The orange and green scattered points represent the experimental value of Pro0 and Pro1, respectively.
   (d) Experimental demonstration for the uncertainty relations of the additive form. The solid magenta, solid violet, and dashed violet curves, in turn, represent the theoretical values of Sum0, Sum1, and Sum2, where $\textrm{Sum0}={\left( {\Delta {S_x}} \right)^2} + {\left( {\Delta {S_y}} \right)^2} + {\left( {\Delta {S_z}} \right)^2}$, $\textrm{Sum1}=\tau \left( {\left| {\langle {S_x}\rangle } \right| + \left| {\langle {S_y}\rangle } \right| + \left| {\langle {S_z}\rangle } \right|} \right)/2$, and $\textrm{Sum2}= \left( {\left| {\langle {S_x}\rangle } \right| + \left| {\langle {S_y}\rangle } \right| + \left| {\langle {S_z}\rangle } \right|} \right)/2$. The magenta and violet scattered points represent the experimental value of Sum0 and Sum1, respectively.
   Error bars represent $\pm1$ s.d. %The error bars are smaller than the data markers in Fig.~(b), but are much longer in Figs.~(c) and (d). The lengthening is due to error propagation.
   %The error bars are smaller than the markers in Fig.~\textbf{b}, and
   }
    \label{latitude}
\end{figure}

The diamond we use is a bulk sample with the $^{13}$C nuclide at the natural abundance of about 1.1\% and the nitrogen impurity less than 5 ppb.
The NV center is composed of one substitutional nitrogen atom and an adjacent vacancy as shown in Fig.~\ref{structure}(a).
The electronic ground state $^3A_2$ is a triplet state and has a zero-field splitting of about 2.87 GHz. With a static magnetic field of around 510 G applied along the NV axis, both the electron spin and the host nitrogen nuclear spin are polarized by optical pumping \cite{Jacques,Sar}.
The two levels $\left| {{m_s} = 0} \right\rangle$ and $\left| {{m_s} =  - 1} \right\rangle$ act as a spin-$1/2$ system or qubit which is manipulated by resonant microwave (MW) pulses.
% and has the dephasing time of about 1.5 $\mu$s.
%The spin-$1/2$ system, encoded as $\left| {{m_s} = 0} \right\rangle  \leftrightarrow \left| {{m_s} =  - 1} \right\rangle$ and correspondingly labeled by $\left| 0 \right\rangle  \leftrightarrow \left| 1 \right\rangle$, is manipulated by resonant microwave (MW) pulses.
The spin state can be read out by optical excitation and red fluorescence detection. To enhance the fluorescence collection, a solid immersion lens is fabricated on the diamond above the NV center \cite{Rong2015,Robledo}.
In the following, the rotating frame determined by the resonant MW is adopted, and in this rotating frame, the two levels $\left| {{m_s} = 0} \right\rangle$ and $\left| {{m_s} =  - 1} \right\rangle$ are labeled by $\left| 0 \right\rangle$ and $\left| 1 \right\rangle$, respectively.

At first the qubit is initialized to the state $\left| 0 \right\rangle$, and then the desired state $\left| \psi \right\rangle$ is prepared by an operation $U$. After that, an operation $V$ is applied. Finally, the measurement in the $\{ \left| 0\right\rangle, \left| 1\right\rangle\}$ basis, or equivalently, the measurement of the observable $S_z$, is performed. The combined effect of the operation $V$ and the measurement of $S_z$ amounts to the measurement of ${V^\dag}S_zV$.
%$(S_x + S_y)/\sqrt 2$, $(S_y + S_z)/\sqrt 2$, and $(S_z + S_x)/\sqrt 2$, can be constructed by adjusting $V$.
In our experiment, the process from initialization to measurement is repeated four million times to acquire the expectation of ${V^\dag }{S_z}V$ associated with the state $\left| \psi \right\rangle$. The standard deviation can then be calculated as $\Delta \left( {{V^\dag }{S_z}V} \right) = \sqrt{1/4-\langle {V^\dag }{S_z}V\rangle^{2}}$. Different observables, including $S_x$, $S_y$, and $S_z$, can be constructed by adjusting $V$.

\begin{figure}\centering
\includegraphics[width=1\columnwidth]{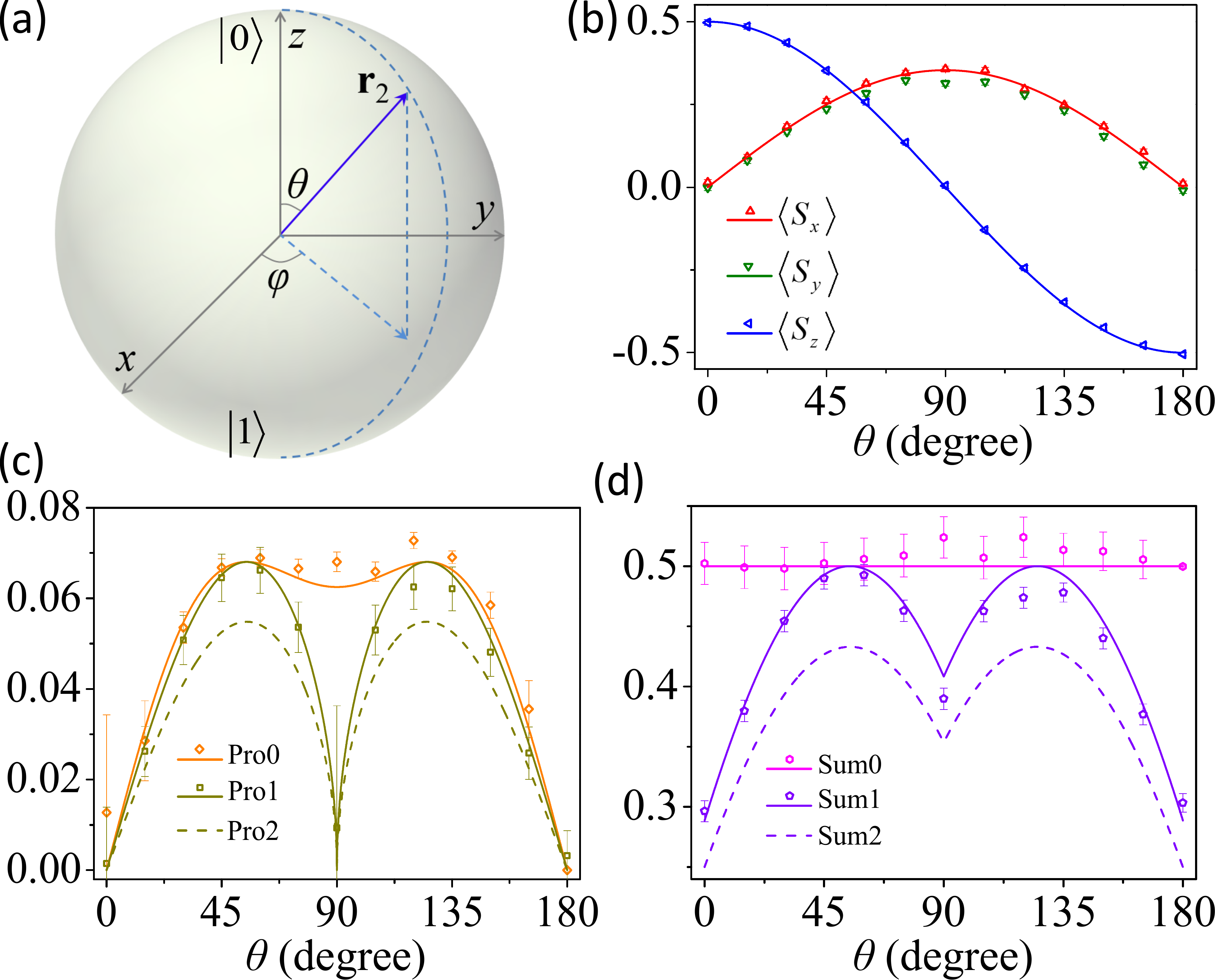}
\caption{Experimental results.
   (a) Bloch sphere and the vector ${\bf{r}}_2$ with $\varphi  = \pi /4$.
   (b) Expectations of $S_x$, $S_y$, and $S_z$.
   (c) Experimental demonstration for the uncertainty relation of the multiplicative form.
   (d) Experimental demonstration for the uncertainty relations of the additive form. All the symbols have the same meaning as those in Fig.~\ref{latitude}.
   }
    \label{meridian}
\end{figure}

We select two series of pure states $\left| \psi \right\rangle$ with Bloch vectors ${\bf{r}}_1 = \left( \sqrt {2/3}\cos \varphi, \sqrt {2/3}\sin \varphi,1/\sqrt 3 \right)$ and ${\bf{r}}_2 = \left( \sin \theta / \sqrt 2,\sin \theta / \sqrt 2,\cos \theta\right)$ as illustrated in Figs.~\ref{latitude}(a) and \ref{meridian}(a). The expectations of $S_x$, $S_y$, and $S_z$ are illustrated in Figs.~\ref{latitude}(b) and \ref{meridian}(b).
The verification of the multiplicative uncertainty relation in (\ref{spin-product}) is shown in Figs.~\ref{latitude}(c) and \ref{meridian}(c). The results demonstrate that for the product $\Delta {S_x} \Delta {S_y} \Delta {S_z}$, the lower bound with the triple constant $\tau$, namely, ${\left| {\tau ^3\langle {S_x}\rangle \langle {S_y}\rangle \langle {S_z}\rangle }/8 \right|^{1/2}}$, outperforms the naive lower bound without the triple constant, namely, ${\left| {\langle {S_x}\rangle \langle {S_y}\rangle \langle {S_z}\rangle }/8 \right|^{1/2}}$.
The results for the sum ${\left( {\Delta {S_x}} \right)^2} + {\left( {\Delta {S_y}} \right)^2} + {\left( {\Delta {S_z}} \right)^2}$ are shown in Figs.~\ref{latitude}(d) and \ref{meridian}(d). The lower bound with the triple constant $\tau$, namely, $\tau \left( {\left| {\langle {S_x}\rangle } \right| + \left| {\langle {S_y}\rangle } \right| + \left| {\langle {S_z}\rangle } \right|} \right)/2$, outperforms the naive one without the triple constant, namely, $\left( {\left| {\langle {S_x}\rangle } \right| + \left| {\langle {S_y}\rangle } \right| + \left| {\langle {S_z}\rangle } \right|} \right)/2$.
The same results also support the uncertainty relation in (\ref{spin-sum-1/2}) which was derived previously \cite{Abbott,Hofmann}.
Therefore, the triple uncertainty relations (\ref{spin-product}), (\ref{spin-sum}), and (\ref{spin-sum-1/2}) are confirmed by the experimental results.
%All the experimental results confirm the triple uncertainty relations (\ref{spin-product}), (\ref{spin-sum}), and (\ref{spin-sum-1/2}).
%The situation for the variance-based lower bound is similar. The variance-based lower bound with the extra coefficient $8/5$ is superior to the naive one without this coefficient.
%All the experimental results confirm the triple uncertainty relations in (\ref{spin-product}), (\ref{spin-sum}), and (\ref{variance}).

\emph{Discussions}.---
The experimental errors mainly come from the imperfection of microwave pulses and the fluctuation of photon counts. The error bars are smaller than the data markers in Figs.~\ref{latitude}(b) and \ref{meridian}(b), but are much larger in Figs.~\ref{latitude}(c)(d) and \ref{meridian}(c)(d). The enlargement of errors is due to error propagation.

In addition to the triple uncertainty relations (\ref{spin-product}), (\ref{spin-sum}), and (\ref{spin-sum-1/2}), some other uncertainty relations exhibit similar behaviour.
%Our approach can also be applied to derive other form of tight uncertainty relations for triple observables.
For instance, from the inequality
$(\Delta A)^{2}+(\Delta B)^{2}\geq[\Delta (A+B)]^{2}/2$ \cite{Maccone}, one has
$$
\begin{aligned}
&{\left( {\Delta {S_x}} \right)^2} + {\left( {\Delta {S_y}} \right)^2} + {\left( {\Delta {S_z}} \right)^2} \ge \\
&\frac{1}{4} \left\{ {{{\left[ {\Delta ({S_x} + {S_y})} \right]}^2} + {{\left[ {\Delta ({S_y} + {S_z})} \right]}^2} + {{\left[ {\Delta ({S_z} + {S_x})} \right]}^2}} \right\}.
\end{aligned}
$$
By tightening the above inequality for spin-$1/2$ we have
\begin{equation}\label{variance}
\begin{aligned}
&{\left( {\Delta {S_x}} \right)^2} + {\left( {\Delta {S_y}} \right)^2} + {\left( {\Delta {S_z}} \right)^2} \ge \frac{2}{5} \times \\
&\left\{ {{{\left[ {\Delta ({S_x} + {S_y})} \right]}^2} + {{\left[ {\Delta ({S_y} + {S_z})} \right]}^2} + {{\left[ {\Delta ({S_z} + {S_x})} \right]}^2}} \right\}.
\end{aligned}
\end{equation}
The equality holds if and only if the spin-$1/2$ system is in a pure state which should also satisfy $r_x + r_y + r_z = 0$.
%This relation is stronger than .

One can also derive tight entropic uncertainty relations for spin-1/2.
The typical entropic uncertainty relation derived by Maassen and Uffink is given by
$H( {\left. A \right|\rho }) + H( {\left. B \right|\rho }) \ge  - 2\log c(A,B)$,
where $H( {\left. A \right|\rho })$ and $H( {\left. B \right|\rho })$ are the Shannon entropy of the measurement outcomes of $A$ and $B$ with a given state $\rho$, and the number
$c(A,B)={\max _{i,j}}\left| {\left\langle {{{a_i}}}
 \mathrel{\left | {\vphantom {{{a_i}} {{b_j}}}}
 \right. \kern-\nulldelimiterspace}
 {{{b_j}}} \right\rangle } \right|$
with $\left| {{a_i}} \right\rangle$ and $\left| {{b_j}} \right\rangle$ being the eigenstates of $A$ and $B$, respectively \cite{Maassen}.
From the pairwise relations
$ H({{S_x}}) + H({{S_y}}) \ge \log 2$, $H({{S_y}}) + H({{S_z}}) \ge \log 2$, and
$ H({{S_z}}) + H({{S_x}}) \ge \log 2$, one has
$H({{S_x}}) + H({{S_y}}) + H({{S_z}}) \ge \frac{3}{2}\log 2$
which is again not tight. A tight lower bound needs an additional factor ${\tau}^2$ \cite{Sanchez93,Sanchez95,Wehner08}, i.e.,
\begin{equation}\label{spin-entropy-log4}
H({{S_x}}) + H({{S_y}}) + H({{S_z}}) \ge \log 4 = \frac{3{\tau ^2}}{2}\log 2.
\end{equation}
The equality holds if and only if the qubit is in one of the eigenstates of $S_x$, $S_y$, or $S_z$.

Additionally, we conjecture that the relation (\ref{spin-product}) is also valid for any spin quantum number $s$. The equality holds when $\langle S_x^2\rangle = \langle S_y^2\rangle = \langle S_z^2\rangle = s(s+1)/3$ and $\left| {\langle {S_x}\rangle } \right|=\left| {\langle {S_y}\rangle } \right|=\left| {\langle {S_z}\rangle } \right|=s/\sqrt3$, and also holds when $\left( {\left| {\langle {S_x}\rangle } \right| - s} \right)\left( {\left| {\langle {S_y}\rangle } \right| - s} \right)\left( {\left| {\langle {S_y}\rangle } \right| - s} \right) = 0$.

\emph{Conclusion}.---
We have derived tight uncertainty relations for the triple components of angular momentum in the spin-$1/2$ representation. The triple constant $\tau$ exhibits its universality to some extent.
%We have derived tight uncertainty inequalities for the preparation uncertainty associated with the triple spin-$1/2$ components.
%The universality and the origin of the triple constant $\tau$ have been revealed.
The experimental demonstration with the single spin of an NV center consistently supports the theoretical results.
Our work enriches the uncertainty relations of more than two observables.
%Our work enriches the product form uncertainty relations of more than two observables.

This work was supported by the 973 Program (Grants No.~2013CB921800 and No.~2016YFA0502400), the National Natural Science Foundation of China (Grants No.~11227901, No.~31470835, No.~11275131, and No.~91636217), the China Postdoctoral Science Foundation (Grant No. 2016M600997), the CAS (Grants No.~XDB01030400 and No.~QYZDY-SSW-SLH004, No.~YIPA2015370), and the Fundamental Research Funds for the Central Universities (WK2340000064).

W. M. and B. C. contributed equally to this work.


\begin{thebibliography}{99}

\bibitem{Heisenberg}
%Heisenberg, W. \"{U}ber den anschaulichen Inhalt der quantentheoretischen Kinematik und Mechanik. \emph{Z. Phys.} \textbf{43}, 172-198 (1927); \emph{Quantum Theory and Measurement}, edited by J. A. Wheeler and W. H. Zurek (Princeton Univ. Press, Princeton, New Jersey, 1983), p.~62.
W. Heisenberg, \"{U}ber den anschaulichen Inhalt der quantentheoretischen Kinematik und Mechanik, Z. Phys. \textbf{43}, 172 (1927); \emph{Quantum Theory and Measurement}, edited by J. A. Wheeler and W. H. Zurek (Princeton Univ. Press, Princeton, New Jersey, 1983), p.~62.

\bibitem{Kennard}
%Kennard, E. H. Zur Quantenmechanik einfacher Bewegungstypen. \emph{Z. Phys.} \textbf{44}, 326-352 (1927).
E. H. Kennard, Zur Quantenmechanik einfacher Bewegungstypen, Z. Phys. \textbf{44}, 326 (1927).

\bibitem{Weyl}
%Weyl, H. Gruppentheorie Und Quantenmechanik (Hirzel, Leipzig, 1928).
H. Weyl, \emph{Gruppentheorie Und Quantenmechanik} (Hirzel, Leipzig, 1928).

\bibitem{Robertson29}
%Robertson, H. P. The uncertainty principle. \emph{Phys. Rev.} 34, 163-164 (1929).
H. P. Robertson, The uncertainty principle, Phys. Rev. \textbf{34}, 163 (1929).

\bibitem{Schroedinger}
%Schr\"{o}dinger, E. Zum Heisenbergschen Unsch\"{a}rfeprinzip. \emph{Sitz. Preuss. Akad. Wiss.} \textbf{14}, 296-303 (1930); A. Angelow and M.-C. Batoni, arXiv: 9903100.
E. Schr\"{o}dinger, Zum Heisenbergschen Unsch\"{a}rfeprinzip, Sitz. Preuss. Akad. Wiss. (Phys.-Math. Klasse) \textbf{19}, 296 (1930); A. Angelow and M.-C. Batoni, About Heisenberg Uncertainty Relation, arXiv: 9903100v3 (2008).



\bibitem{Busch07}
%Busch, P., Heinonen, T. ${\text{\&}}$ Lahti, P. Heisenberg¡¯s uncertainty principle. \emph{Phys. Rep.} \textbf{452}, 155-176 (2007).
P. Busch, T. Heinonen, and P. Lahti, Heisenberg¡¯s uncertainty principle, Phys. Rep. \textbf{452}, 155 (2007).

\bibitem{Busch13PRL}
%Busch, P., Lahti, P. ${\text{\&}}$ Werner, R. F. Proof of Heisenberg¡¯s Error-Disturbance Relation. \emph{Phys. Rev. Lett.} \textbf{111}, 160405 (2013).
P. Busch, P. Lahti, and R. F. Werner, Proof of Heisenberg¡¯s Error-Disturbance Relation, Phys. Rev. Lett. \textbf{111}, 160405 (2013).

\bibitem{Busch14PRA}
%Busch, P., Lahti, P. ${\text{\&}}$ Werner, R. F. Heisenberg uncertainty for qubit measurements. \emph{Phys. Rev. A} \textbf{89}, 012129 (2014).
P. Busch, P. Lahti, and R. F. Werner, Heisenberg uncertainty for qubit measurements, Phys. Rev. A \textbf{89}, 012129 (2014).

\bibitem{Busch14RMP}
%Busch, P., Lahti, P. ${\text{\&}}$ Werner, R. F. Quantum root-mean-square error and measurement uncertainty relations. \emph{Rev. Mod. Phys.} \textbf{86}, 1261-1281 (2014).
P. Busch, P. Lahti, and R. F. Werner, Quantum root-mean-square error and measurement uncertainty relations, Rev. Mod. Phys. \textbf{86}, 1261 (2014).

\bibitem{Buscemi}
%Buscemi, F., Hall, M. J. W., Ozawa, M. ${\text{\&}}$ Wilde, M. M. Noise and disturbance in quantum measurements: an information-theoretic approach. \emph{Phys. Rev. Lett.} \textbf{112}, 050401 (2014).
F. Buscemi, M. J. W. Hall, M. Ozawa, and M. M. Wilde, Noise and disturbance in quantum measurements: an information-theoretic approach, Phys. Rev. Lett. \textbf{112}, 050401 (2014).

\bibitem{Sulyok}
%Sulyok, G. \emph{et al}. Experimental test of entropic noise-disturbance uncertainty relations for spin-1/2 measurements. \emph{Phys. Rev. Lett.} \textbf{115}, 030401 (2015).
G. Sulyok, S. Sponar, B. Demirel, F. Buscemi, M. J. W. Hall, M. Ozawa, and Y. Hasegawa, Experimental test of entropic noise-disturbance uncertainty relations for spin-1/2 measurements, Phys. Rev. Lett. \textbf{115}, 030401 (2015).

%\bibitem{Rastegin}
%A. E. Rastegin, Uncertainty and Certainty Relations for Successive Projective Measurements of a Qubit in Terms of Tsallis' Entropies. Commun. Theor. Phys. \textbf{63}, 687 (2015).

\bibitem{Ma}
%Ma, W. \emph{et al}. Experimental demonstration of Heisenberg's measurement uncertainty relation based on statistical distances. \emph{Phys. Rev. Lett.} {\bf 116} (2016) 160405.
W. Ma, Z. Ma, H. Wang, Y. Liu, Z. Chen, F. Kong, Z. Li, M. Shi, F. Shi, S.-M. Fei, and J. Du, Experimental demonstration of Heisenberg's measurement uncertainty relation based on statistical distances, Phys. Rev. Lett. \textbf{116} 160405 (2016).

\bibitem{Hirschman}
%Hirschman, I. I. A note on entropy. \emph{Am. J. Math.} \textbf{79}, 152-156 (1957)
I. I. Hirschman, A note on entropy, Am. J. Math. \textbf{79}, 152 (1957).

\bibitem{Beckner}
%Beckner, W. Inequalities in Fourier analysis. \emph{Ann. Math.} \textbf{102}, 159-182 (1975).
W. Beckner, Inequalities in Fourier analysis, Ann. Math. \textbf{102}, 159 (1975).

\bibitem{Birula75}
%Bia\l{}ynicki-Birula, I. ${\text{\&}}$ Mycielski, J. Uncertainty relations for information entropy in wave mechanics. \emph{Commun. Math. Phys.} \textbf{44}, 129-132 (1975).
I. Bia\l{}ynicki-Birula and J. Mycielski, Uncertainty relations for information entropy in wave mechanics, Commun. Math. Phys. \textbf{44}, 129 (1975).

\bibitem{Deutsch}
%Deutsch, D. Uncertainty in Quantum Measurements. \emph{Phys. Rev. Lett.} \textbf{50}, 631-633 (1983).
D. Deutsch, Uncertainty in Quantum Measurements, Phys. Rev. Lett. \textbf{50}, 631 (1983).

\bibitem{Kraus}
%Kraus, K. Complementary observables and uncertainty relations. \emph{Phys. Rev. D} \textbf{35}, 3070-3075 (1987).
K. Kraus, Complementary observables and uncertainty relations, Phys. Rev. D \textbf{35}, 3070 (1987).

\bibitem{Maassen}
%Maassen, H. ${\text{\&}}$  Uffink, J. B. M. Generalized entropic uncertainty relations. \emph{Phys. Rev. Lett.} \textbf{60}, 1103-1106 (1988).
H. Maassen and J. B. M. Uffink, Generalized entropic uncertainty relations, Phys. Rev. Lett. \textbf{60}, 1103 (1988).

\bibitem{Braunstein}
S. L. Braunstein, C. M. Caves, and G. J. Milburn, Generalized Uncertainty Relations: Theory, Examples, and Lorentz Invariance, Ann. Phys. \textbf{247}, 135 (1996).

\bibitem{Sanchez98}
J. S\'{a}nchez-Ruiz, Optimal entropic uncertainty relation in two-dimensional Hilbert space, Phys. Lett. A \textbf{244}, 189 (1998).

\bibitem{Ghirardi}
G.C. Ghirardi, L. Marinatto, R. Romano, Optimal entropic uncertainty relation in two-dimensional Hilbert space, Phys. Lett. A \textbf{317}, 32 (2003).

\bibitem{Birula06}
%Bialynicki-Birula, I. Formulation of the uncertainty relations in terms of the R\'{e}nyi entropies. \emph{Phys. Rev. A} \textbf{74}, 052101 (2006).
I. Bialynicki-Birula, Formulation of the uncertainty relations in terms of the R\'{e}nyi entropies, Phys. Rev. A \textbf{74}, 052101 (2006).

\bibitem{Wehner10}
%Wehner, S. ${\text{\&}}$ Winter, A. Entropic uncertainty relations---a survey. \emph{New J. Phys.} \textbf{12}, 025009 (2010).
S. Wehner and A. Winter, Entropic uncertainty relations---a survey, New J. Phys. \textbf{12}, 025009 (2010).

\bibitem{Birula11}
%Bialynicki-Birula, I. \& Rudnicki, {\L}. Entropic Uncertainty Relations in Quantum Physics. \emph{Statistical Complexity} 1-34 (Springer, Netherlands, 2011).
I. Bialynicki-Birula and {\L}. Rudnicki, Entropic Uncertainty Relations in Quantum Physics, \emph{Statistical Complexity}, edited by K. D. Sen (Springer, Dordrecht, 2011), p. 1.

\bibitem{Partovi}
%Partovi, M. H. Majorization formulation of uncertainty in quantum mechanics. \emph{Phys. Rev. A} \textbf{84}, 052117 (2011).
M. H. Partovi, Majorization formulation of uncertainty in quantum mechanics, Phys. Rev. A \textbf{84}, 052117 (2011).

\bibitem{Puchala}
%Pucha{\l}a, Z., Rudnicki, {\L} \& Zyczkowski, K. Majorization entropic uncertainty relations. {\it J. Phys. A: Math. Theor.} \textbf{46}, 272002 (2013).
Z. Pucha{\l}a, {\L}. Rudnicki, and K. Zyczkowski, Majorization entropic uncertainty relations, J. Phys. A: Math. Theor. \textbf{46}, 272002 (2013).

\bibitem{Friedland}
%Friedland, S., Gheorghiu, V. \& Gour, G. Universal Uncertainty Relations. {\it Phys. Rev. Lett.} \textbf{111}, 230401 (2013).
S. Friedland, V. Gheorghiu, and G. Gour, Universal Uncertainty Relations, Phys. Rev. Lett. \textbf{111}, 230401 (2013).

\bibitem{Maccone}
%Maccone, L. ${\text{\&}}$ Pati, A. K. Stronger Uncertainty Relations for All Incompatible Observables. \emph{Phys. Rev. Lett.} \textbf{113}, 260401 (2014).
L. Maccone and A. K. Pati, Stronger Uncertainty Relations for All Incompatible Observables, Phys. Rev. Lett. \textbf{113}, 260401 (2014).

\bibitem{Zhang}
%Zhang, J., Zhang, Y. \& Yu, C. S. R\'{e}nyi entropy uncertainty relation for successive projective measurements. {\it Quant. Inform. Processing} \textbf{14}, 2239 (2015).
J. Zhang, Y. Zhang, and C.-s. Yu, R\'{e}nyi entropy uncertainty relation for successive projective measurements, Quant. Inform. Processing \textbf{14}, 2239 (2015).

\bibitem{Li}
J.-L. Li, C.-F. Qiao, Reformulating the Quantum Uncertainty Relation, Sci. Rep. \textbf{5}, 12708 (2015).

\bibitem{Coles}
P. J. Coles, M. Berta, M. Tomamichel, S. Wehner, Entropic Uncertainty Relations and their Applications, arXiv: 1511.04857 (2015).

\bibitem{Robertson34}
%Robertson, H. P. An Indeterminacy Relation for Several Observables and Its Classical Interpretation. \emph{Phys. Rev.} \textbf{46}, 794-801 (1934).
H. P. Robertson, An Indeterminacy Relation for Several Observables and Its Classical Interpretation, Phys. Rev. \textbf{46}, 794 (1934).

\bibitem{Ivanovic}
I. D. Ivanovic, An inequality for the sum of entropies of unbiased quantum measurements, J. Phys. A: Math. Gen. \textbf{25}, 363 (1992).

\bibitem{Sanchez93}
J. S\'{a}nchez, Entropic uncertainty and certainty relations for complementary observables, Phys. Lett. A \textbf{173}, 233 (1993).

\bibitem{Sanchez95}
J. S\'{a}nchez-Ruiz, Improved bounds in the entropic uncertainty and certainty relations for complementary observables, Phys. Lett. A \textbf{201}, 125 (1995).

\bibitem{Trifonov}
%Trifonov, D. A. ${\text{\&}}$ Donev, S. G. Characteristic uncertainty relations. \emph{J. Phys. A: Math. Gen.} \textbf{31}, 8041-8047 (1998).
D. A. Trifonov and S. G. Donev, Characteristic uncertainty relations, J. Phys. A: Math. Gen. \textbf{31}, 8041 (1998).

\bibitem{Shirokov}
%Shirokov, M. I. Interpretation of Uncertainty Relations for Three or More Observables. arXiv: quant-ph/0404165 (2004).
M. I. Shirokov, Interpretation of Uncertainty Relations for Three or More Observables, arXiv: 0404165 (2004).

\bibitem{Pati}
%Pati, A. K. \& Sahu, P. K. Sum uncertainty relation in quantum theory. \emph{Phys. Lett. A} \textbf{367}, 177-181 (2007).
A. K. Pati and P. K. Sahu, Sum uncertainty relation in quantum theory, Phys. Lett. A \textbf{367}, 177 (2007).

\bibitem{Wehner08}
S. Wehner and A. Winter, Higher entropic uncertainty relations for anti-commuting observables, J. Math. Phys. \textbf{49}, 062105 (2008).

\bibitem{Huang}
%Huang, Y. Variance-based uncertainty relations. \emph{Phys. Rev. A} \textbf{86}, 024101 (2012).
Y. Huang, Variance-based uncertainty relations, Phys. Rev. A \textbf{86}, 024101 (2012).

\bibitem{Kaniewski}
J. Kaniewski, M. Tomamichel, and S. Wehner, Entropic uncertainty from effective anticommutators, Phys. Rev. A \textbf{90}, 012332 (2014).

\bibitem{Chen}
%Chen, B. ${\text{\&}}$ Fei, S.-M. Sum uncertainty relations for arbitrary $N$ incompatible observables. \emph{Sci. Rep.} \textbf{5}, 14238 (2015).
B. Chen and S.-M. Fei, Sum uncertainty relations for arbitrary $N$ incompatible observables, Sci. Rep. \textbf{5}, 14238 (2015).

\bibitem{Abbott}
A. A. Abbott, P.-L. Alzieu, M. J. W. Hall, and C. Branciard, Tight State-Independent Uncertainty Relations for Qubits, Mathematics \textbf{4}, 8 (2016).

\bibitem{Chen2016-1}
%Chen, B., Fei, S.-M. ${\text{\&}}$ Long, G.-L. Sum uncertainty relations based on Wigner¨CYanase skew information. \emph{Quantum Inf. Process.} \textbf{15}, 2639-2648 (2016).
B. Chen, S.-M. Fei, and G.-L. Long, Sum uncertainty relations based on Wigner-Yanase skew information, Quantum Inf. Process. \textbf{15}, 2639 (2016).

\bibitem{Chen2016-2}
%Chen, B., Cao, N.-P, Fei, S.-M. ${\text{\&}}$ Long, G.-L. Variance-based uncertainty relations for incompatible observables. \emph{Quantum Inf. Process.} \textbf{15}, 3909-3917 (2016).
B. Chen, N.-P. Cao, S.-M. Fei, and G.-L. Long, Variance-based uncertainty relations for incompatible observables, Quantum Inf. Process. \textbf{15}, 3909 (2016).

\bibitem{Weigert}
%Kechrimparis, S. ${\text{\&}}$ Weigert, S. Heisenberg uncertainty relation for three canonical observables. \emph{Phys. Rev. A} \textbf{90}, 062118 (2014).
S. Kechrimparis and S. Weigert, Heisenberg uncertainty relation for three canonical observables, Phys. Rev. A \textbf{90}, 062118 (2014).

\bibitem{Dammeier}
%Dammeier, L., Schwonnek, R. ${\text{\&}}$ Werner, R. F. Uncertainty relations for angular momentum. \emph{New J. Phys.} \textbf{17}, 093046 (2015).
L. Dammeier, R. Schwonnek, and R. F. Werner, Uncertainty relations for angular momentum, New J. Phys. \textbf{17}, 093046 (2015).

\bibitem{sm}
See Supplemental Material at ..., which includes Ref.~\cite{Rong2014}, for theoretical and experimental deltails.

\bibitem{Rong2014}
X. Rong, J. Geng, Z. Wang, Q. Zhang, C. Ju, F. Shi, C.-K. Duan, and J. Du, Implementation of Dynamically Corrected Gates on a Single Electron Spin in Diamond, Phys. Rev. Lett. \textbf{112}, 099903 (2014).

\bibitem{Hofmann}
%Hofmann, H. F. ${\text{\&}}$ Takeuchi, S. Violation of local uncertainty relations as a signature of entanglement. \emph{Phys. Rev. A} \textbf{68}, 032103 (2003).
H. F. Hofmann and S. Takeuchi, Violation of local uncertainty relations as a signature of entanglement, Phys. Rev. A \textbf{68}, 032103 (2003).

%\bibitem{Gruber}
%A. Gruber, A. Dr\"{a}benstedt, C. Tietz, L. Fleury, J. Wrachtrup, and C. von Borczyskowski, Scanning Confocal Optical Microscopy and Magnetic Resonance on Single Defect Centers, Science \textbf{276}, 2012 (1997).

\bibitem{Doherty}
M. W. Doherty, N. B. Manson, P. Delaney, F. Jelezko, J. Wrachtrup, L. C. L. Hollenberg, The nitrogen-vacancy colour centre in diamond, Phys. Rep. \textbf{528}, 1 (2013).

\bibitem{Schirhagl}
R. Schirhagl, K. Chang, M. Loretz, and C. L. Degen, Nitrogen-Vacancy Centers in Diamond: Nanoscale Sensors for Physics and Biology, Annu. Rev. Phys. Chem. \textbf{65}, 83 (2014).

\bibitem{Prawer}
S. Prawer and I. Aharonovich, \emph{Quantum Information Processing with Diamond} (Woodhead, Cambridge, England, 2014).

\bibitem{Wrachtrup}
J. Wrachtrup and A. Finkler, Single spin magnetic resonance, J. Magn. Reson. \textbf{269}, 225 (2016).

% long coherence time
%\bibitem{Balasubramanian2009}
%G. Balasubramanian, P. Neumann, D. Twitchen, M. Markham, R. Kolesov, N. Mizuochi, J. Isoya, J. Achard, J. Beck, J. Tissler, V. Jacques, P. R. Hemmer, F. Jelezko, and J. Wrachtrup, Ultralong spin coherence time in isotopically engineered diamond, Nat. Mater. \textbf{8}, 383 (2009).
%
%\bibitem{Maurer}
%P. C. Maurer, G. Kucsko, C. Latta, L. Jiang, N. Y. Yao, S. D. Bennett, F. Pastawski, D. Hunger, N. Chisholm, M. Markham, D. J. Twitchen, J. I. Cirac, and M. D. Lukin, Room-Temperature Quantum Bit Memory Exceeding One Second, Science \textbf{336}, 1283 (2012).

% high fidelity
%\bibitem{Rong2014}
%X. Rong, J. Geng, Z. Wang, Q. Zhang, C. Ju, F. Shi, C.-K. Duan, and J. Du, Implementation of Dynamically Corrected Gates on a Single Electron Spin in Diamond, Phys. Rev. Lett. \textbf{112}, 099903 (2014).

% sensing
%\bibitem{Maze}
%J. R. Maze, P. L. Stanwix, J. S. Hodges, S. Hong, J. M. Taylor, P. Cappellaro, L. Jiang, M. V. Gurudev Dutt, E. Togan, A. S. Zibrov, A. Yacoby, R. L. Walsworth, and M. D. Lukin, Nanoscale magnetic sensing with an individual electronic spin in diamond, Nature (London) \textbf{455}, 644 (2008).

%\bibitem{Balasubramanian2008}
%G. Balasubramanian, I. Y. Chan, R. Kolesov, M. Al-Hmoud, J. Tisler, C. Shin, C. Kim, A. Wojcik, P. R. Hemmer, A. Krueger, T. Hanke, A. Leitenstorfer, R. Bratschitsch, F. Jelezko, and J. Wrachtrup, Nanoscale imaging magnetometry with diamond spins under ambient conditions, Nature (London) \textbf{455}, 648 (2008).
%
%\bibitem{Grinolds}
%M. S. Grinolds, S. Hong, P. Maletinsky, L. Luan, M. D. Lukin, R. L. Walsworth, and A. Yacoby, Nanoscale magnetic imaging of a single electron spin under ambient conditions, Nat. Phys. \textbf{9}, 215 (2013).
%
%\bibitem{Muller}
%C. M\"{u}ller, X. Kong, J.-M. Cai, K. Melentijevi¡äc, A. Stacey, M. Markham, D. Twitchen, J. Isoya, S. Pezzagna, J.Meijer, J. Du, M. B. Plenio, B. Naydenov, L. P. McGuinness, and F. Jelezko, Nuclear magnetic resonance spectroscopy with single spin sensitivity, Nat. Commun. \textbf{5}, 4703 (2014).
%
%\bibitem{Shi2015}
%F. Shi, Q. Zhang, P. Wang, H. Sun, J. Wang, X. Rong, M. Chen, C. Ju, F. Reinhard, H. Chen, J. Wrachtrup, J. Wang, and J. Du, Science 347, 1135 (2015).
%
%\bibitem{Jakobi}
%I. Jakobi, P. Neumann, Y. Wang, D. D. Bhaktavatsala Rao, F. El Hallak, M. A. Bashir, M. Markham, A. Edmonds, D. Twitchen, and J. Wrachtrup, Measuring broadband magnetic fields on the nanoscale using a hybrid quantum register, Nat. Nanotech. \textbf{12}, 67 (2017).

% quantum metrology on NV
%\bibitem{Waldherr2012}
%G. Waldherr, J. Beck, P. Neumann, R. S. Said, M. Nitsche, M. L. Markham, D. J. Twitchen, J. Twamley, F. Jelezko, and J. Wrachtrup, High-dynamic-range magnetometry with a single nuclear spin in diamond, Nat. Nanotech. \textbf{7}, 105 (2012).
%
%\bibitem{Liu}
%G.-Q. Liu, Y.-R. Zhang, Y.-C. Chang, J.-D. Yue, H. Fan, X.-Y. Pan, Demonstration of entanglement-enhanced phase estimation in solid, Nat. Commun. \textbf{6}, 6726 (2015).


% quantum computation on NV
%\bibitem{Shi2010}
%F. Shi, X. Rong, N. Xu, Y. Wang, J. Wu, B. Chong, X. Peng, J. Kniepert, R.-S. Schoenfeld, W. Harneit, M. Feng, and J. Du, Room-Temperature Implementation of the Deutsch-Jozsa Algorithm with a Single Electronic Spin in Diamond, Phys. Rev. Lett. \textbf{105}, 040504 (2010).


\bibitem{Jacques}
V. Jacques, P. Neumann, J. Beck, M. Markham, D. Twitchen, J. Meijer, F. Kaiser, G. Balasubramanian, F. Jelezko, and J. Wrachtrup, Dynamic Polarization of Single Nuclear Spins by Optical Pumping of Nitrogen-Vacancy Color Centers in Diamond at Room Temperature, Phys. Rev. Lett. \textbf{102}, 057403 (2009).

\bibitem{Sar}
T. van der Sar,	Z. H. Wang,	M. S. Blok,	H. Bernien,	T. H. Taminiau,	D. M. Toyli, D. A. Lidar, D. D. Awschalom, R. Hanson, and V. V. Dobrovitski, Decoherence-protected quantum gates for a hybrid solid-state spin register, Nature (London) \textbf{484}, 82 (2012).

%\bibitem{Waldherr2014}
%G. Waldherr, Y. Wang, S. Zaiser, M. Jamali, T. Schulte-Herbr¨¹ggen, H. Abe, T. Ohshima, J. Isoya, J. Du, P. Neumann, and J. Wrachtrup, Quantum error correction in a solid-state hybrid spin register, Nature \textbf{506}, 204 (2014).
%
%\bibitem{Zu}
%C. Zu, W.-B. Wang, L. He, W.-G. Zhang, C.-Y. Dai, F. Wang, and L.-M. Duan, Experimental realization of universal geometric quantum gates with solid-state spins, Nature \textbf{514}, 72 (2014).
%
%\bibitem{Camejo}
%S. Arroyo-Camejo, A. Lazariev, S. W. Hell, and G. Balasubramanian, Room temperature high-fidelity holonomic single-qubit gate on a solid-state spin, Nat. Commun. \textbf{5}, 4870 (2014).
%
%\bibitem{Kong}
%F. Kong, C. Ju, P. Huang, P. Wang, X. Kong, F. Shi, L. Jiang, and J. Du, Experimental Realization of High-Efficiency Counterfactual Computation, Phys. Rev. Lett. \textbf{115}, 080501 (2015).
%
%\bibitem{Xu}
%K. Xu, T. Xie, Z. Li, X. Xu, M. Wang, X. Ye, F. Kong, J. Geng, C. Duan, F. Shi, and J. Du, Experimental Adiabatic Quantum Factorization under Ambient Conditions Based on a Solid-State Single Spin System, Phys. Rev. Lett. \textbf{118}, 130504 (2017).

% fundamental physics on NV
%\bibitem{Huang2011}
%P. Huang, X. Kong, N. Zhao, F. Shi, P. Wang, X. Rong, R.-B. Liu, J. Du, Observation of an anomalous decoherence effect in a quantum bath at room temperature, Nat. Commun. \textbf{2}, 570 (2011).
%
%\bibitem{Waldherr2011}
%G. Waldherr, P. Neumann, S. F. Huelga, F. Jelezko, and J. Wrachtrup, Violation of a Temporal Bell Inequality for Single Spins in a Diamond Defect Center, Phys. Rev. Lett. \textbf{107}, 090401 (2011).
%
%\bibitem{Reinhard}
%F. Reinhard, F. Shi, N. Zhao, F. Rempp, B. Naydenov, J. Meijer, L. T. Hall, L. Hollenberg, J. Du, R.-B. Liu, and J. Wrachtrup, Tuning a Spin Bath through the Quantum-Classical Transition, Phys. Rev. Lett. \textbf{108}, 2004029 (2012).
%
%\bibitem{Zhou}
%J. Zhou, P. Huang, Q. Zhang, Z. Wang, T. Tan, X. Xu, F. Shi, X. Rong, S. Ashhab, and J. Du, Observation of Time-Domain Rabi Oscillations in the Landau-Zener Regime with a Single Electronic Spin, Phys. Rev. Lett. \textbf{112}, 010503 (2014).
%
%\bibitem{Hensen}
%B. Hensen, H. Bernien, A. E. Dreau, A. Reiserer, N. Kalb, M. S. Blok, J. Ruitenberg, R. F. L. Vermeulen, R. N. Schouten, C. Abellan, W. Amaya, V. Pruneri, M. W. Mitchell, M. Markham, D. J. Twitchen, D. Elkouss, S. Wehner, T. H. Taminiau, and R. Hanson, Loophole-free Bell inequality violation using electron spins separated by 1.3 kilometres, Nature (London) \textbf{526}, 682 (2015).

\bibitem{Robledo}
L. Robledo, L. Childress, H. Bernien, B. Hensen, P. F. A. Alkemade, and R. Hanson, High-fidelity projective read-out of a solid-state spin quantum register, Nature \textbf{477}, 574 (2011).

\bibitem{Rong2015}
X. Rong, J. Geng, F. Shi, Y. Liu, K. Xu, W. Ma, F. Kong, Z. Jiang, Y. Wu, and J. Du, Experimental fault-tolerant universal quantum gates with solid-state spins under ambient conditions, Nat. Commun. \textbf{6}, 8748 (2015).

\end{thebibliography}
\end{document}